# Magnetic field dependence of the electrical conductivity in $V_3Ge$ in the vicinity of $H_{c2}$

V. A. Marchenko and A. V. Nikulov

*Institute of Solid State Physics, USSR Academy of Sciences*

The excess electrical conductivity of $V_3Ge$ mono-crystal has been investigated experimentally in a perpendicular magnetic field near the second critical field $H_{c2}$. At fields $H > 0.97 H_{c2}$ the experimental data agree with theoretical calculations for the paraconductivity in the Hartree approximation. At $H = 0.97 H_{c2}$ the conductivity increases sharply and the critical current appears. This sharp qualitative change is attributed to a transition to the Abrikosov state.

It is assumed[1] that in type-II superconductors a second-order phase transition from the normal state to a mixed state (the Abrikosov state) occurs at the second critical field $H_{c2}$. However, this transition has not been studied sufficiently. Moreover, in measurements of the specific heat it was shown[2] that near $H_{c2}$ the experimental results are described well by a one-dimensional model and a $\lambda$ anomaly characteristic of any second-order phase transition has not been observed. Since a phase transition cannot occur in one-dimensional systems, we should ask where does the normal state become the Abrikosov vortex lattice which has been observed many times in fields lower than $H_{c2}$.

In order to investigate this problem we have measured the excess electrical conductivity $\Delta\sigma = (\sigma - \sigma_N)$ of high-quality $V_3Ge$ samples with very weak pinning as a function of the magnetic field at $T = 4.2 K$. The usual four-point method was used in the measurements. The magnetic field was directed perpendicularly to the long axis of the sample, along which the electric current flowed. The conductivity in the magnetic field $H$ is defined as $\sigma = I/V(l/S)$, where $I$ is the current flowing through the sample, $V$ is the voltage between the potential contacts in the magnetic field $H$, $l$ is the distance between the potential contacts ($\approx$3-5 mm), $S$ is the cross-sectional area of the sample ($\approx 0.5$ mm$^2$), and $\sigma_N$ is the conductivity of the sample in the normal state (as $H \to \infty$). The other characteristics of the sample and the experimental conditions are given in Ref. 3.

The measurement results are shown in Fig. 1. The theoretical dependence for paraconductivity (i.e. the excess electrical conductivity of the normal state caused by fluctuation pairing of electrons) is obtained in the Hartree approximation from the linear Ami-Maki theory,[4] by using Eq. (10) from Ref. 5, in which it was shown that a superconductor can be regarded as a one-dimensional system near $H_{c2}$. In our case $N(0) = 9.73 \times 10^{41} J^{-1} cm^{-3}$ (Ref. 6), $\xi(0) = 48 A$, $T_{c0} = 6 K$, $T = 4.2 K$, $h = 0.293$ and $H_{c2}$, which is not a singular point, play the role of an adjustment parameter. The $\Delta\sigma/\sigma_N(H/H_{c2})$ dependence, Fig.1, can be divided into three regions $H/H_{c2} > 0.973$, $0.969 < H/H_{c2} < 0.973$, and $H/H_{c2} < 0.969$. In the first region the critical current is equal to zero and the conductivity does not depend on the current, which is characteristic of the normal state; the experimental data agree with the Hartree approximation for the paraconductivity. In the second region there is a sharp increase of the excess conductivity, and the critical current appears. It is interesting to note that the observed sharp increase in the excess conductivity occurs against a background of a smooth variation due to fluctuation pairing of electrons. In the third region the critical current has a finite value and the conductivity depends on the current, which is characteristic of the Abrikosov vortex state; the experimental data do not agree with the Hartree approximation for the paraconductivity.

The above-described dependence of $\Delta\sigma$ on the magnetic field can be interpreted in the following manner: the agreement between the experimental and calculated values means that the superconductor is in a one-dimensional fluctuation state in the first region. In the second region, a transition occurs from the one-dimensional state to the Abrikosov state. We believe that the difference between the transition field (0.97 $H_{c2}$) and $H_{c2}$ is attributable to the fact that the entropy term in the Ginzburg-Landau "free energy" has not been completely taken into account, i.e., the Ginzburg-Landau "free energy" is essentially a block Hamiltonian.[7]

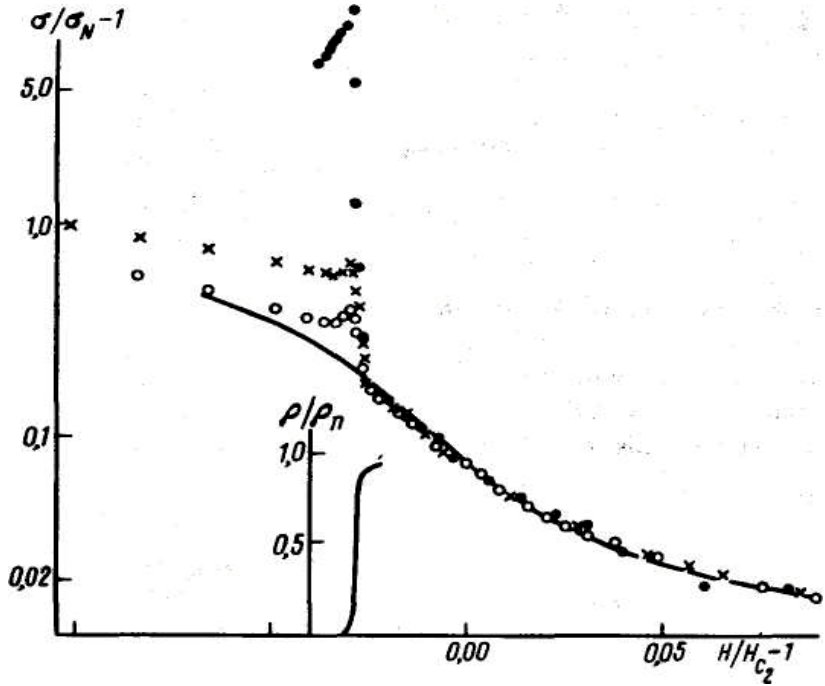

FIG. 1. Magnetic-field dependence of the excess conductivity for different measuring currents through the sample: •-$j$ = 0.5 A/cm$^2$, ×-$j$ = 2 A/cm$^2$, and o-$j$ =10 A/cm$^2$; — is the theoretical dependence. The magnetic-field dependence of the resistance is shown at the bottom of the figure for $j$= 0.05 A/cm$^2$. $j=I/S$ is the density of the electric current flowing through the sample.

It is interesting to note that the width of the transition region (*0.004 $H_{c2}$*) is much narrower than that of the specific heat "jump" at $H_{c2}$ that has been observed in superconductors with similar $\xi(0)$ value under similar conditions (identical $t=T/T_{c0}$).[5] It is possible that in the temperature dependence of the specific heat in the vicinity of $H_{c2}$ there is a narrow anomaly, in addition to the broad "jump", which could not be observed in the experiment.

In conclusion, the authors thank V. Zavaritskii for reading the manuscript and for his comments.